\def\rf#1;#2;#3;#4;#5 {\par#1, {\it #3} {\bf #4}, #5 (#2). \par}

\def\beq#1{\begin{equation}\label{#1}}
\def\eeq{\end{equation}}
\def\beqa#1{\begin{eqnarray}\label{#1}}
\def\eeqa{\end{eqnarray}}

\def\ignore#1{}

\def\spose#1{\hbox to 0pt{#1\hss}}
\def\simlt{\mathrel{\spose{\lower 3pt\hbox{$\mathchar"218$}}
     \raise 2.0pt\hbox{$\mathchar"13C$}}}
\def\simgt{\mathrel{\spose{\lower 3pt\hbox{$\mathchar"218$}}
     \raise 2.0pt\hbox{$\mathchar"13E$}}}
\def\simpropto{\mathrel{\spose{\lower 3pt\hbox{$\mathchar"218$}}
     \raise 2.0pt\hbox{$\propto$}}}

\def\ed{\end{document}}
\def\bl{\vskip0.423truecm}
\def\ind{\noindent\hskip1.8truecm}
\def\Section#1{{\raggedright\bl\bl\goodbreak{\noindent\bf#1}\bl}}

\documentstyle[11pt]{article}

\begin{document}
\baselineskip=0.423truecm
\def\basel{\normalsize\baselineskip=0.423truecm}


{\hfill}
\vskip1.222truecm


\noindent
{\bf 
IS THE UNIVERSE REALLY SO SIMPLE?
}
  
\bl\bl\bl
\ind
Milan M. \'Cirkovi\'c

\bl\ind 
{\it Astronomical Observatory Belgrade

}

\ind 
{\it Volgina 7, 11160 Belgrade}

\ind   
{\it Serbia, YUGOSLAVIA}

\ind   
{\it arioch@eunet.yu}

\bl\bl\bl\ind
Received August 13, 2001; revised September 02, 2001

\bl\bl

\noindent
The intriguing recent suggestion of Tegmark that the universe---contrary to all our 
experiences and expectations---contains only a small amount of information due to an 
extremely high degree of internal symmetry is critically examined. It is shown that 
there are several physical processes, notably Hawking evaporation of black holes and 
non-zero decoherence time effects described by Plaga, as well as thought experiments 
of Deutsch and Tegmark himself, which can be construed as arguments against the 
low-information universe hypothesis. Some ramifications for both quantum mechanics 
and cosmology are briefly discussed. 
\bl


\clearpage

\Section{1. INTRODUCTION: LIU HYPOTHESIS}

\noindent
1.1.	The quantity of information present in our universe and its transfer has been an intriguing issue in theoretical cosmology ever since pioneering studies of Rindler$^{(1)}$ and Metzner and Morrison.$^{(2)}$ In 1970-ies and early 1980-ies it has gained a new momentum after seminal results of Bekenstein$^{(3)}$ and Hawking$^{(4)}$ on the intricate relationship between information, thermodynamics and gravitation. Finally, during the last decade several new moments and motivations have been put forward. Treumann has explicated the usage of information-theoretical methods in cosmology and demonstrated the crucial role of inflation in bringing about observed information richness of the present universe.$^{(5)}$ The great advances in the nascent discipline of physical eschatology$^{(6)}$ motivated investigations into possibilities of future information processing, tightly connected with the issue of future survival of intelligent observers.$^{(7)}$ In addition, from the point of view of both classical and quantum information theory, as well as general computer science, this link between quantum mechanics and cosmology is highly interesting.$^{(8,9,10)}$ Therefore, it seems that applications of information theory and theory of complexity in cosmology has gradually reached mature methodological state. In addition, the explosive growth of computer science and industry in recent years has stimulated many interesting discussions on the possibility of large-scale simulations of human sensory inputs and natural surroundings.$^{(11)}$ 

1.2.	In this vein, by far the most far-reaching and provocative hypothesis has been put forward by Tegmark,$^{(12)}$ which we shall call the low-information universe (henceforth LIU) picture. Tegmark's paper is rich and fascinating reading, many ideas of which are rather difficult to convey in a review manner. In brief, the main idea relies on conjuction of the Everett's "no-collapse" quantum theory and properties of temporal evolution of nonlinear systems. According to the "no-collapse" theory, all components of the linear superposition of the universal wave function exist simultaneously.$^{(13)}$ This view, coupled with our understanding of the structure formation in the universe and non-linear amplification of small perturbations, gives rise to the following thought-provoking picture. Initially very simple (= of low information content) universe evolves in such way that the quantity of algorithmic information present in it is constant throughout the cosmological evolution. Therefore, the total amount of algorithmic information remains at the same low level as it was at the beginning of structure formation processes. However, due to the constant symmetry breaking realized through branching of Everett's wavefunction components, apparent complexity increases without limits, since only one component of the "grand superposition" is accessible to any inside observer (the latter fact is explained by quantum mechanical decoherence). In Tegmark's words, "The net result is that although the wavefunction of the universe contains almost no algorithmic information (we can specify it by simply giving the initial wave function and the Hamiltonian), and will retain for instance translational and rotational symmetry from the initial data, we will experience an asymmetric universe that appears extremely complex."

1.3.	It is important to understand that Tegmark's hypothesis is a new physical theory, and not just an "interpretation" or a "view". Not only does it tell us something new (and very unorthodox) about the nature of the physical reality, but it makes several distinct and well-defined predictions which could ultimately be falsified. We shall consider some of these physical predictions below. Although it is not possible to falsify any of them at the present moment, we claim that simultaneous satisfaction of different empirical constraints makes the theory less probable. In any case, we ought to avoid the trap (partially of semantic nature) which is easy to fall into, as testified by the example of Everett's multiverse theory, to claim that LIU is just another "interpretation" of the same underlying theory, which can not be meaningfully discriminated from other such "interpretations".\footnote{There is a slowly building consensus that the Everett's "no-collapse" view is a separate theory, not just an interpretation of the quantum mechanical formalism.$^{(14,15,16,17)}$ Some of the (still thought) experiments discriminating between Everett's theory and the orthodox quantum mechanics are described by Deutsch$^{(18)}$ and Plaga.$^{(19)}$ More on this in Section 3.2. below.}

1.4.	Having this in mind, in the further course of this essay we shall try to elaborate and defend the following four theses: 								

{\bf (A)} LIU does not preclude the existence of definite quantity of the apparent information in our universe. The latter is still interesting quantity for physics. 			

{\bf (B)} LIU relies on the idealization of complete separation of multiverse branches, or zero decoherence time. However, the effect recently proposed for experimental verification by Plaga may obviate this assumption. Even more remote verification proposals, like those of Deutsch, Page and Tegmark conceptually jeopardize the theory. 						

{\bf (C)} The well-known possibility of solving the information loss paradox in black hole evaporation by fundamental non-unitarity (Hawking's idea) is incompatible with LIU. 									

{\bf (D)} Intelligent observers can not create any new information beyond the pre-existent linear superposition. Therefore, LIU entails an extreme version of physical reductionism.

1.5.	One thing should be kept in mind: LIU hypothesis cannot be brought down by "frontal assault", i.e.\ through finding some empirical phenomena whose complexity is so high that they cannot be incorporated in LIU scheme. The reason is obvious enough, as Tegmark points out: any such phenomenon would be a {\it local\/} one, and its complicated description is subsumed in the overall symmetry. (In the same manner as any one particular book in Borges' "Library of 
Babel"---having a very long description---is subsumed by a simple general rule, like "All combinations of letters of the specified (book) length actually exist in the Library."$^{(37)}$) Thus we shall concentrate upon those aspects of the physical mechanism envisaged by Tegmark, like linearity or unitarity, on which the process of subsuming depends. In addition, we shall reflect upon issues, like epistemic status of the apparent information or reductionism, which may not have cosmological relevance, but may impact our manner of doing physics if the LIU hypothesis is a correct description of reality.

1.6.	Our discussion intends to show that, with no further empirical information available, the probability of LIU being the true description of reality is significantly diminished by these additional considerations.

\Section{2. SPEAKABLE AND UNSPEAKABLE IN THE COSMOLOGICAL EVOLUTION}

\noindent

2.1.	 One of the main merits of Tegmark's paper is its conceptual clarity. The 
author emphasizes: "We have argued that although the universe appears to contain an 
enormous, perhaps even infinite, amount of information, this impression is nonetheless consistent with the assumption that its algorithmic information content is quite small, 
even if no new physics whatsoever is invoked. In short, reality could in fact be much simpler than it appears, with its apparent plethora or complex structures such as trees, fountains and rabbits… Thus even if both the laws of physics and the initial conditions exhibit so much symmetry that they are easy to describe, we would not expect most of these symmetries to be manifested in what we see when we open our eyes." (Ref. 12, p. 36). However, the true issue is whether science in general and physics in particular deals with algorithmic or only apparent information. (We shall call the information actually perceived in any single slice of the Everett's multiverse the {\it apparent information}, in accordance with Tegmark's terminology. It is obviously a complementary concept to the {\it algorithmic information}, as used by Tegmark and other authors.) It is, unfortunately, well known that there is a recognized vagueness regarding the exact meaning and proper usage of the concept of information, resulting in some authors even calling the usage of this term "vulgar".$^{(20)}$ In the present study, we claim that only apparent information is physically relevant quantity (although this does not warrant that it is well defined in all conceivable physical contexts!). In other words, it is the only "speakable" issue in the physical world, in contradistinction to the Platonic world of mathematical objects (to borrow a picture from Penrose, Ref. 21). One of the difficulties with the LIU concept is that it subtly shifts grounds between the physical world and the abstract mathematical description; this is partly responsible for shocking effect of its main idea.

2.2.	 When we talk about apparent informational content of any one particular slice in the universal superposition, we should note that it is not a completely exact physical concept, since no empirical data really pertain to a single well-defined slice, due to the continuous branching of the universal wavefunction. Thus, any set of data describes not a single slice, but a family of what we can call {\it close slices}, i.e.\ those produced by "recent" and "local" measurement-like interactions. In spite of this rather technical difficulty, one may argue that empirical data and theoretical models pertaining to the informational content of the apparent universe (such as the results of Treumann$^{(5)}$) are good approximations to the informational content of a single component of the universal superposition. Since natural laws are valid in any one single branch of the multiverse, its informational content evolves according to the expectations for the complex systems dominated by long-range forces, mainly gravity.

2.3.	 One may conclude that even on the face of LIU hypothesis, cosmological evolution (a single history or small set of close histories) is an entirely tractable problem of interest for cosmologists. Methods, such as those used by Treumann,$^{(5)}$ and general application of approximations such as Bekenstein-Hawking formula for entropy of gravitating systems (e.g.\ Ref.\ 21), are part of the "speakable" reality even if the totality of everything that exists is trivially simple.

2.4.	 In a sense, this issue has been recognized by Tegmark, as may be inferred from his locutions, such as "that only macroscopically 'classical' states could be perceived by self-aware subsets of the universe (such as us)" in which he talks about observers. It has been left unsaid, but it is an obvious consequence of LIU that out of all branches of the universal wavefunction we may expect most of them to be empty as far as observers ("self-aware elements") are concerned. In particular, the process of nonlinear structure formation is essentially fine-tuned in order to produce galaxies (and consequently stars, planets, etc.) since the spectrum of primordial perturbations might as well produce conditions in which either there is no decoupling of initial overdensities from the Hubble flow, or the amplitude of perturbations is so high that matter ends up predominantly in supermassive black holes. In all these cases, there would be no conditions for development of life and subsequently, self-awareness. This anthropic selection effect related to the degree of symmetry in initial conditions for structure formation has been considered long ago by many authors.$^{(15,21,22,23)}$ Thus, even the complete preservation of the grand symmetry with nonlinear evolution apparently violates the principle of sufficient reason by introducing an epistemological asymmetry through emergence of only a small subset of the universal wavefunction components containing observers. In other words, the distribution of the {\it apparent\/} complexity will nevertheless be wildly non-uniform (in spite of what could be {\it prima facie\/} expected in case of such an extreme symmetry). This tendency is unidirectional in cosmic time.\footnote{The same circumstance goes some steps toward mitigating Tegmark's (pseudo)Copernican suggestion that the LIU hypothesis is somehow further diminishing humanity's place in the universe; more on this below.}

2.5.	 Finally, some would argue that this issue is mainly a semantical one. This is false, since the question "Which type of information is data X-Y-Z that some observer perceives?" is substantially different from-and in a sense subordinate to-the question "Which type of information the observer may sensibly talk about (as perceived or not)?" The latter question is only implicitly treated in Tegmark's paper, but it is an important complement to the issues raised in that study.

\Section{3. ISOLATION OR DESOLATION?}

\noindent
3.1. Tegmark's idea crucially depends on separate Everett's branches being completely and absolutely sealed off at all times.\footnote{For the sake of definiteness, we may take a single branch as the referent one, and state that at any instant of time measured by the ideal clock in that particular branch, off-diagonal terms in the density matrix are zero.} It is easy to understand qualitatively why this is so: an enormous amount of algorithmic information could be encoded in cross-correlation terms (expressing post-interaction correlations between different components in the global superposition). It is much more difficult to give a quantitative estimate of the magnitude of this effect, especially since it strongly depends on the chosen theory of measurement. Thus, one of the predictions of LIU hypothesis is that-within what one may call the many-world paradigm-any theories of dynamical state reduction, similar to the ones developed by Ghirardi, Rimini and Weber$^{(24)}$ or Di\`osi$^{(25)}$ within the single-world context, which include nonlinear terms in the Schr\"odinger's equation, will be unavoidably falsified by experiments. This may apply, for instance, to the toy model of nonlinear quantum theory suggested by Weinberg$^{(26)}$ in order to test validity of the EPR-type experiments. As explained by Polchinski,$^{(27)}$ Weinberg's model leads to what he has dubbed "Everett phone," i.e.\ the possibility of inter-world communication (see other instances below). Any such theory, however, would be rather contrived by definition, since, as Tegmark himself explains, the mechanism of environmental decoherence is sufficient for "simulation" of any measurement-like effects observed so far. Potentially, one may expect that in the course of further development of such theories, the extra amount of information coded in these non-linear terms could be exactly calculated, which will show whether there are any versions of such theories compatible with the LIU hypothesis. In absence of such calculations, the LIU requirement does seem a stringent constraint. This occurs notwithstanding the somewhat strange nature of the argument, reminscent of Hoyle-Narlikar cosmological model of 1960-ies and 1970-ies, arguing the necessity of changes in local microphysics on the basis of special cosmological boundary conditions required.$^{(28)}$ On the other hand, this prediction demonstrates again that the LIU hypothesis is not an arbitrary "interpretation", but a legitimate and testable physical theory.

3.2. Similarly, any conceivable situation in which communication with other branches can be established is detrimental to the LIU hypothesis. This applies, among other instances, to the experiment recently suggested by Plaga,$^{(19)}$ as well as the older thought-experiment of Deutsch.$^{(18)}$ In the first case, Plaga introduces a "gateway state" between the two hypothetical worlds created as the outcome of a conventional photon polarization measurement. The gateway is a microscopic part of the apparatus which is isolated sufficiently, so that its decoherence timescale is long, and thus it "sees" the global superposition long enough to be influenced by it. In a detailed technical account, Plaga suggests that sufficiently isolated ions in electromagnetic traps can be used as indicators of such influences, if the experimental setup is ingenious enough, and the instruction given to experimenters can be carried out with sufficient precision. In the Deutsch's {\it Gedankenexperiment}, after two histories decohere in a conventional manner, they are {\it re-cohered\/} by a convenient and ingenious manipulation of the relevant Hamiltonians in post-measurement time. In Deutsch's own words, the net result of the experiment is an anomalous lack of correlation (if Everett is right and there is no wavefunction collapse): "The interference phenomenon seen by our observer at the end of the experiment requires the presence of the both spin values, though he accurately remembers having known at a previous time that only one of them was present. He must infer that there was more than one copy of himself (and the atom) in existence at that time, and that these copies merged to form his present self." (Ref.\ 18, p.\ 37) The difference between these two examples is that, while Plaga's experiment considers immediate post-measurement interaction of the two wavefunction branches through a gateway state (in fact, it is crucial for the claimed feasibility of this experiment to have an extremely fast measurement procedure), the one of Deutsch deals with post-measurement interaction after an arbitrarily long time. The price payed for the latter advantage is "only" the necessity of having an operational quantum computer capable of simulating human-level intelligence.\footnote{As far as the no-collapse view is concerned, the nature of the computer is in fact irrelevant, since it presupposes quantum mechanics as "the universal theory" (the very title of Deutsch's article), and therefore {\it any\/} working computer is already part of the quantum world. If we stick to somewhat 
more cautious epistemological stance, we should emphasize that the observer in this thought experiment must be quantum in nature, {\it and\/} it is crucial that the Hamiltonian 
expressing his/her internal "self-interaction(s)" is known. In principle, the advances in biophysics might bring about the complete knowledge of microscopic processes within a biological observer (like human brain), as well as the relevant technology to modify such processes in order to intentionally induce necessary changes in the internal Hamiltonian. (Among various other features, this thought experiment thus clearly demonstrates ontological realism inherent in Everett's theory.) However, it seems more realistic that this degree of knowledge and manipulative powers will be reached by a human-made quantum computer.}  
Thus, the Deutsch experiment belongs to not-so-near future as far as technology is concerned, but it is conceptually important, since it shows {\it demonstrable\/} difference between "collapse" and "no-collapse" versions of quantum mechanics. {\it A fortiori}, LIU is a testable scientific hypothesis, as already argued above. In addition, it enables us to-at least in principle-experimentally test the very {\it nature of information\/} science is dealing with, no mean achievement by any set of standards. 

3.3. If one feels, with Polchinski,$^{(27)}$ that "communication between branches of the wave function seems even more bizarre than faster-than-light communication and consequent loss of Lorentz invariance",\footnote{"...but it is not clear that it represents an actual inconsistency." (Ref.\ 27, p.\ 399).}  one may invoke the quantum suicide (thought?) experiment proposed by Tegmark.$^{(16)}$ In this scenario, a quantum measurement results in firing of the gun pointed to the head of the experimenter if one particular value of, for instance, spin {\it z}-projection of a fermion is measured, and in harmless "click" of the gun at the other outcome. Notice that only physical collapse---as in the "orthodox" Copenhagen interpretation or the dynamical reduction theories---is actually harmful from the experimenter's point of view. Since in the "no-collapse" view there is no actual collapse, just fast decoherence between the branches, experimenter will find herself in the strange situation of impossibility of committing suicide, although the gun is loaded and fully functional! This is different from the "outsider" view of, say, assistant in the experiment, who will perceive the bloody deed after at most several repetitions of the experiment (being in one of the decohered branches). This experiment may indirectly support the LIU hypothesis by showing that the quantum mechanical collapse does not happen. However, it remains an open and difficult epistemological issue whether this "I-know-but-cannot-tell" type of experiment may discriminate between various theories. A contrary opinion may be heard, based on the statement that the question "what will you perceive" is not a well-defined question in a quantum mechanical context (Prof. Ken Olum, private communication). We cannot enter into discussion of this issue here.

3.4. There may be a loophole left to Tegmark to defend the LIU theory in the following form. Even if the evolution of quantum mechanical systems is fundamentally nonlinear, one may expect the algorithmic information content of the multiverse density matrix today would be almost the same as for the initial one as long as the equation governing its evolution is deterministic. And determinism is invoked here in its "strongest" form, i.e.\ similar to the Laplacian case: one neglects the apparent "indeterminism" resulting from observer's subjective incapacity of perceiving the entire picture. Whether we should expect possible nonlinear terms in the Schr\"odinger evolution to be deterministic or stochastic, remains a fully open question. The already mentioned Ghirardi et al.\ theory$^{(24)}$ has been argued to be intrinsically indeterministic.$^{(29)}$ However, one thing is clear: the determinism necessary for LIU to work is---in the field of philosophy of time---compatible only with B-theories of time (no temporal becoming). Whether B-theories are superior to A-theories (entailing reality of temporal becoming), or even tenable is an age-old problem not being solved to this day. Recently, however, Elitzur and Dolev$^{(30)}$ have forcefully argued---contrary to the prevailing opinion so far---that A-theories are actually closer to the physical ideas, as far as quantum gravitational effects are concerned. This leads us directly to the substance of the thesis {\bf (C)}.

\Section{4. LOSS OF INFORMATION = LOSS OF SIMPLICITY}


4.1. The algorithmic information once present in the universe does not change during the amplification of small perturbations in the highly symmetric reality as envisaged by LIU. Conventionally, one tends to see this as a counterargument to the idea of information {\it growth\/} in the course of the cosmological evolution. However, the idea that the amount of algorithmic information may be irretrievably {\it lost\/} in the course of evolution of a physical system is equally antithetical to LIU. Any form of non-conservation of information cannot be accomodated in the LIU framework. Of course, the information loss must be fundamental, not just apparent from the human point of view; the stock example of waves erasing text written in sand on a beach does not suffice for the task, since the microscopic evolution of the entire (sea + beach) system is conventionally assumed to be unitary, i.e. the information contained in the text is preserved on microscopic level for all eternity.

4.2. It is well known that the Hawking process of black hole evaporation is claimed to violate this rule, for the following reason. Let us consider a pure quantum state corresponding to a distribution of matter of mass $M \gg M_{Pl}$ (Planck mass), which collapses under its own weight. The density matrix of such state is given as $\rho = |\psi \rangle \langle \psi |$, with vanishing entropy $S = - {\rm Tr} (\rho \ln \rho)$. If $M$ is high enough, the matter will inevitably form a black hole. Subsequently, the black hole will slowly evaporate by Hawking process, emitting blackbody radiation (which by definition carries out no information). The semiclassical treatment used by Hawking in discovery of the black hole evaporation and in all subsequent discussions will certainly break down when the mass of black hole approaches MPl, but what will happen with the information from the initial state still locked in the black hole? This is the puzzle of black hole information loss.$^{(20,31)}$ As is well-known, the possibility Hawking himself proposed is that black hole simply evaporates completely and the information is irreversibly lost.$^{(32)}$ Although this idea remains the simplest and the least problematic answer to the puzzle, it has provoked a lot of controversy, since it implies that the evolution of the complete system (universe + black hole) is fundamentally non-unitary, and leads to evolution of pure into mixed quantum states.  In contradistinction to the example above, we can quote the familiar example of the {\it Encyclopaedia Britannica\/} thrown into a black hole. Its informational content is lost forever, if Hawking's idea about nonunutarity is correct.\footnote{Alternatively, it may be lost only from {\it our\/} universe, if the proposal Giddings ascribes to Freeman Dyson,$^{(31)}$ or the fascinating cosmological ideas of Smolin$^{(33)}$ are correct. In these views, there are one or more new universes created inside any black hole, to which the collapsed information is transmitted, without loss of unitarity on the global ("multiverse") scale. The multiverse here is not the multiverse of quantum mechanics ("Everett's multiverse" = the totality of wavefunction branches), but the multiverse of quantum cosmology ("Linde's multiverse" = set of different cosmological domains, presumably causally and/or topologically disconnected from the domain of ours). Obviously, these proposals are highly speculative, to say at least.}

4.3. Thus, the LIU predictions for the solution of the information-loss puzzle are that either information is actually radiated via the higher-order effects in the Hawking radiation (solution proposed, among others, by t'Hooft$^{(34)}$), or that stable remnants of the order of Planck mass remain, encoding an arbitrarily high amount of information.$^{(31)}$ Of course, the information spoken of in both cases is just the apparent one. The two cases are different {\it sensu stricto}, since the first option (information transmitted to the universe) necessarily invokes further symmetry breaking and subsequent creation of an enormous number of "new" histories, which is not necessarily so in the second option (sufficiently isolated black hole may leave a remnant without any measurement-like interaction). However, although the informational content is only apparent, the prediction is physical enough to be falsified. Apart from support it gives LIU as a scientific hypothesis, it should be mentioned that both these options (emission of information or its encoding in remnants; we can classify the possible creation of new universes as a highly special case of remnant solution) possess their own rather formidable internal problems.

4.4. The case of remnants as storage space for information becomes even lesser plausible if some predictions of the currently widely debated M-theories turn out to be correct. It has been proposed, in order to solve the infamous hierarchy problem in particle physics, that new extra dimensions enable significant lowering of the Planck mass scale, as compared to the convention 4-D case.$^{(35,36)}$ It is intuitively clear why lowering of the Planck mass is detrimental to the remnant proposal for explanation of the information loss puzzle: smaller the black hole remnants are, the ratio of information to mass which needs encoding becomes larger, and the entire picture less plausible. In the extreme versions of M-theory proposal, $M_{Pl} \sim 1$ TeV, and the encoding efficiency rises for an astounding factor of  $\sim 10^{16}$. If these ideas receive further confirmation, they would make the case for information conservation even less palatable.

\Section{5. WERE YOUR FEELINGS IN THE INITIAL POWER SPECTRUM?}

\noindent
5.1. The conjecture {\bf (D)} amounts to the idea that initial density perturbations contained, among other things, every single bit of information in a painting of Bruegel, sentence of Plato, a song of Beatles or the shape of any single printed letter in Tegmark's paper. These things are not, however, contained in any "extractable" manner, but are present by virtue of high symmetry and thus are similar to the contents of books in Borges'  "The Library of Babel", in which all conceivable combinations of letters do contain the entire fund of human (and any other expressible) knowledge, which is perfectly useless to any mortal user.$^{(37)}$ However, once a particular set of apparent information has been selected, its temporal evolution is completely determined for all times. In the same manner, the evolution of any particular piece of (apparent) information can be traced back to the spectrum of initial perturbations, and can be effectively {\it reduced\/} to it.

5.2. It is important to understand that LIU theory goes much further than the usual ("moderate") reductionism in this respect. It is, for instance, substantially stronger requirement than that embodied in the usual working philosophy of computationalism,\footnote{A doctrine in the philosophy of mind and cognitive sciences stating (roughly) that cognition and consciousness are instances of the execution of sufficiently complex computational algorithms.$^{(39)}$ Of course, the complexity in this context is just a necessary condition: the search for the {\it right kind\/} of complexity continues.}  since it does not only postulate that all examples of mental phenomena and their consequences mentioned above are algorithmically describable, but also that such algorithms have been pre-existent even to the existence of humankind in the universe! Computationalism {\it per se\/} does not include any specific concept of time and temporal becoming. However, with LIU we loose any freedom in imagining various ways of functioning of human (and other) minds {\it in time}. It is unnecessary to emphasize that any arguments quoted against computationalism in the literature {\it eo ipso\/} count as arguments against the LIU hypothesis.$^{(21,38)}$

5.3. Although this cannot as yet be argued against on purely physical grounds, we may notice that the history of scientific and philosophical thought gives more than one example of failure of such extremely reductionist ideas, the example of ether being probably the most investigated and the best-known one. Other relatively recent cases of failures of reductionist programmes in cosmology include Eddington's numerological "fundamental theory", Bondi-Lyttleton electrical universe,$^{(40)}$ as well as the classical steady state theory of Bondi, Gold and Hoyle.$^{(41)}$ Of course, there have been cases of successful reductionist theories in the history of physics; however, in weighting this issue, one should keep in mind that {\it cosmological theories\/} (such as LIU) in particular should be heavily scrutinized, since it seems that in cosmology reductionistic approach is on a less stable ground than in the other physical disciplines. Furthermore, when judging LIU at the present state of relative ignorance as to the details of cosmological evidence as well as on the empirical merits of the "no-collapse" versions of quantum mechanics, the circumstance that this theory goes much further in reductionism than the ether theory (or any other theory in the history of science!) has to be taken into account. It is important to bear in mind that LIU hypothesis is also antithetical to those versions of the "no-collapse" theory---such as the "many minds" theory$^{(42)}$---which include some non-reductionist elements. 

\Section{6. NOT SO SIMPLE AFTER ALL?}

\noindent
6.1. We have considered several additional issues stemming from Tegmark's LIU theory. The good news for cosmologists interested in application of information and complexity theory to the observable slice of the quantum grand superposition is that dynamical laws will still offer a lot of interesting material to study. If we reject the extreme essentialist position that the goal of science is revealing the "deep" and "true" nature of reality, than the only quantity of true interest to sciences is the apparent information, which is not going to be exhausted in foreseeable future.\footnote{Admittedly bizarre, we may notice that if we {\it do stick\/} to this extreme view, then the LIU hypothesis---if not falsified---represents a true "theory of everything", since apart from those few bytes of information on the initial quantum state of the universe, there is nothing more to be learnt whatsoever. However uninspiring, this prospect may deserve consideration within the framework of epistemological discussions on the topic of status and power of possible "theories of everything" currently under development. It offers rather novel look at the things: we may be able to reach this "holy grail" of physics, but it may possibly be a complete anticlimax---a formally correct, but empirically shallow theory.}  In that respect, consequences of LIU for the astrophysical {\it Weltanschaung\/} are not so serious as it might look at first. The issue of quantity and evolution of apparent informational content of any particular slice or the close related slices remains a legitimate target of physical discourse. Moreover, the dynamics of entropy increase (that is, the manner of structure formation) remains fully meaningful and legitimate physical question. The same applies for sciences other than physical, since they are also, if the LIU theory is correct, just extensions of our knowledge of non-linear, apparent-information-creating processes. The same circumstance does go far toward mitigating Tegmark's dramatic conclusion portraying LIU as "arguing that even the intricate structures that are the subjects of our thoughts, dreams and efforts in life, everything from our loved ones to our parking tickets, are perhaps more aptly described as illusions, as manifestations of the fact that our minds experience the grandeur of reality merely from an extremely limited frog perspective… In short, reality would be much more banal than it appears to be." While it will certainly always remain at least partially a matter of personal taste, in the opinion of the present author, the fact that the (apparent) information content of each component of the superposition is well-defined warrants the label of "good news". The LIU theory, if confirmed, certainly will offer a great leap forward in our understanding of reality, and if falsified, will reveal a host of new insights in the process. This is partially recognized by Tegmark, when he writes: "If we thus restrict our value judgments to empirical considerations, the picture put forward here would have quite positive implications for our ability to do science in the future." However, the restriction to "empirical considerations" seems unnecessary (at least because if restricted in such a manner we would have hardly been able to come up with a theory like LIU!). Perceiving the very deepest symmetry that may exist could as well invoke positive value judgements as to the achievements of science and our ability to do it well in the future. 

6.2. Of course, a further note to be taken is that the narrowness can be construed as a virtue of LIU hypothesis. It is well-known that epistemological criteria favor theories with well-defined predictions over those with predictions based on any number of (more or less free) parameters. A paradigmatic example of this kind in cosmology is to be found in the great controversy between the classical steady-state theory and the relativistic ("Big Bang") world models.$^{(43)}$ It is exactly because of this controversy and of the challenge posed by the rival steady-state paradigm, that the relativistic cosmology became a proper science, with well-defined methodological and epistemic apparatus, and a set of observational criteria for rejection or acceptance of cosmological conjectures. The challenge was crucial exactly because the steady-state was such a narrow, well-defined theory with clear-cut and obviously falsifiable predictions. A similar situation occurs with the LIU hypothesis: a conceptually simple observational falsification of it (say through an unambiguous observation of the final Hawking explosion of black holes together with the proof that extra information can not be hidden in higher orders of the emitted radiation, or a proof that transition of pure into mixed states is hidden somewhere else in nature, say in quantum gravity or in functioning of human brains) will enormously benefit our total cosmological understanding, and may even set a new paradigm on the informational content of everything that exists.

6.3. We conclude that additional requirements {\it a posteriori\/} diminish the probability of the low-information universe hypothesis---fascinating as it is---being the correct description of everything that exists.

\vspace{1cm}
{\bf Acknowledgements.} It is a great pleasure to thank people with whom the author has had thoroughly enjoyable discussions of some of the ideas presented in this essay, especially Prof.\ Petar Gruji\'c, Prof.\ Ken Olum, Ana Vlajkovi\'c, Prof.\ Branislav K.\ Nikoli\'c, Slobodan Popovi\'c and Ivana Dragi\'cevi\'c. Invaluable technical help has been received from Dr.\ Aleksandar B.\ Nedeljkovi\'c, Srdjan Samurovi\'c, Vesna Milo\v sevi\'c-Zdjelar and Nikola Bo\v zinovi\'c. Three anonymous referees are acknowledged for suggestions useful in improving the overall quality of the manuscript. 



\def\refname{\basel{\bf REFERENCES}}

\end{document}